\documentclass[twocolumn,aps,prl,preprintnumbers]{revtex4}
\usepackage{}
\usepackage{bbm}
\usepackage{pifont}
\usepackage[latin9]{inputenc}
\usepackage{amsmath}
\usepackage{amssymb}
\usepackage{graphicx}
\usepackage{epstopdf}
\usepackage{pifont}
\usepackage{mathrsfs}
\usepackage{amsfonts}
\usepackage{amsthm}
\usepackage{color}
\usepackage{txfonts}
\usepackage[colorlinks=true,citecolor=blue,linkcolor=blue,urlcolor=blue,anchorcolor=blue]{hyperref}%
\hypersetup{colorlinks=true,citecolor=blue,linkcolor=blue,urlcolor=blue}
\providecommand{\U}[1]{\protect\rule{.1in}{.1in}}
\makeatletter
\@ifundefined{textcolor}{}
{
\definecolor{BLACK}{gray}{0}
\definecolor{WHITE}{gray}{1}
\definecolor{RED}{rgb}{1,0,0}
\definecolor{GREEN}{rgb}{0,1,0}
\definecolor{BLUE}{rgb}{0,0,1}
\definecolor{CYAN}{cmyk}{1,0,0,0}
\definecolor{MAGENTA}{cmyk}{0,1,0,0}
\definecolor{YELLOW}{cmyk}{0,0,1,0}
}
\makeatother
\begin{document}
\title{Experimental observation of edge-dependent quantum pseudospin Hall effect}
\author{Huanhuan Yang$^{1}$}
\author{Lingling Song$^{1}$}
\author{Yunshan Cao$^{1}$}
\author{X. R. Wang$^{2}$}
\email[Corresponding author: ]{phxwan@ust.hk}
\author{Peng Yan$^{1}$}
\email[Corresponding author: ]{yan@uestc.edu.cn}
\affiliation{$^{1}$School of Electronic Science and Engineering and State Key Laboratory of Electronic Thin Films and Integrated Devices, University of Electronic Science and Technology of China, Chengdu 610054, China}
\affiliation{$^{2}$Physics Department, The Hong Kong University of Science and Technology,
 Clear Water Bay, Kowloon, Hong Kong}

\begin{abstract}
It is a conventional wisdom that the helical edge states of quantum spin Hall (QSH) insulator are particularly stable due to the topological protection of time-reversal symmetry. Here, we report the first experimental observation of an edge-dependent quantum (pseudo-)spin Hall effect by employing two Kekul\'{e} electric circuits with molecule-zigzag and partially-bearded edges, where the chirality of the circulating current in the unit cell mimics the electron spin. We observe a helicity flipping of the topological in-gap modes emerging in opposite parameter regions for the two edge geometries. Experimental findings are interpreted in terms of the mirror winding number defined in the unit cell, the choice of which exclusively depends on the edge shape. Our work offers a deeper understanding of the boundary effect on the QSH phase, and pave the way for studying the spin-dependent topological physics in electric circuits.
\end{abstract}

\maketitle
A paradigm in the topological band insulator family \cite{Hasan2010,Qi2011} is the quantum spin Hall (QSH) insulator, which has an insulating gap in the bulk, but supports gapless helical states on the boundary \cite{Kane2005,Kane20052,Bernevig2006,Konig2007}. QSH insulators are characterized by the topological $\mathbb{Z}_2$ invariant, defined in the presence of time-reversal symmetry. Because of the symmetry protection, the helical edge states are robust against the electronic backscattering \cite{Chen2018,Bernevig2006,Konig2007,Konig2013,Roushan2009}, ushering in a new era in spintronics and quantum computing \cite{Roth2009,Brune2012,Hart2014,Wu2018}. Counterintuitively, Freeney \emph{et al.} recently reported an edge-dependent topology in artificial Kekul\'{e} lattices \cite{Freeney2020}. The mechanism is that the edge geometries of samples determine the choice of the unit cell, and further dictate the value of topological invariants \cite{Fu2011,Slager2013,Kariyado2017,Cao2017,LeeNL2018}. However, the experimental evidence of an edge-dependent quantum (pseduo-)spin Hall effect is still lacking.

\begin{figure*}
  \centering
  \includegraphics[width=0.9\textwidth]{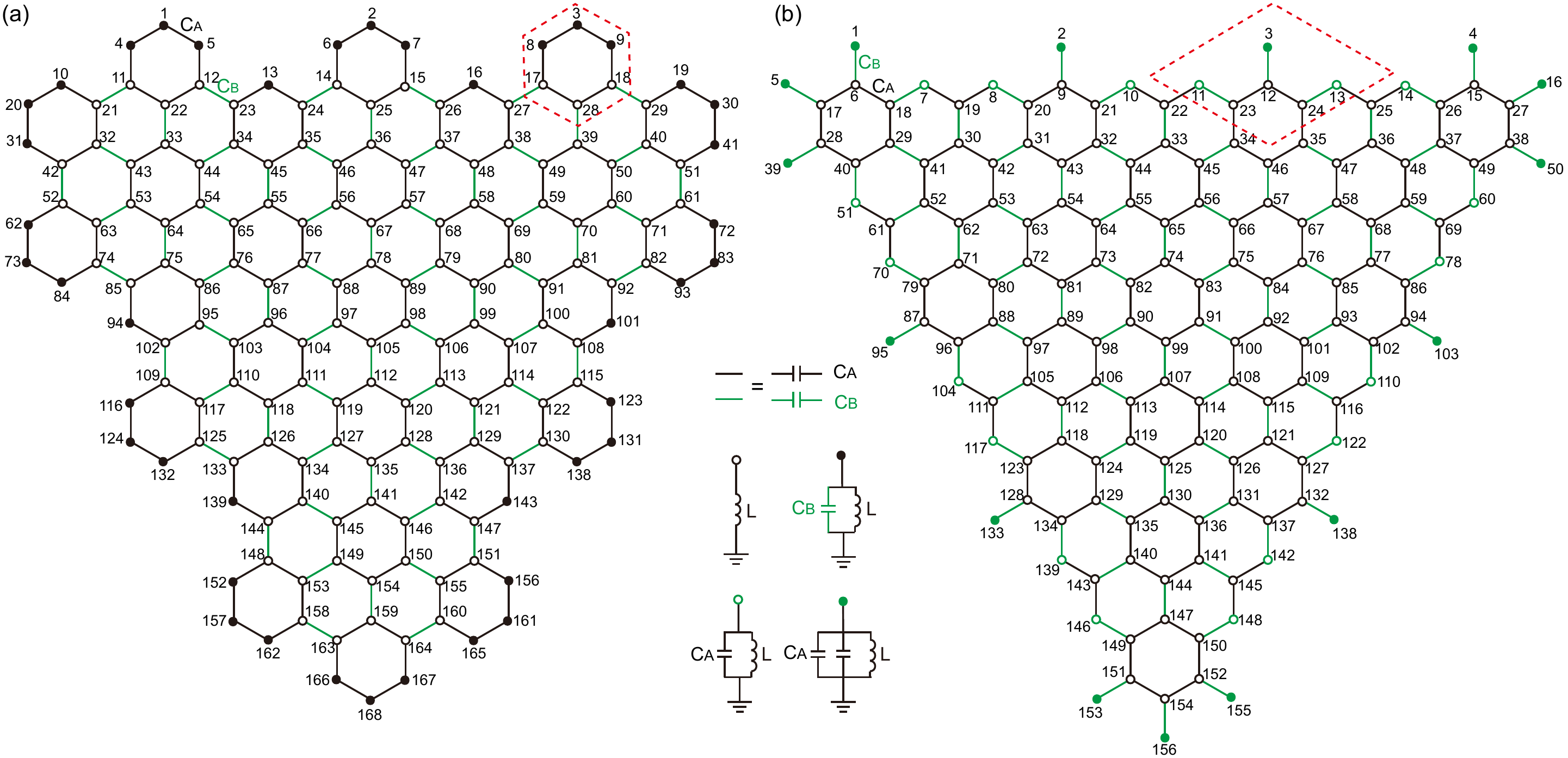}\\
  \caption{Illustration of two artificial Kekul\'{e} LC circuits with (a) molecule-zigzag and (b) partially-bearded edge terminations. Each node is grounded by inductors and capacitors with the configuration shown in the inset. Dashed red hexagon and rhombus represent the approximate unit cells for the two different edge shapes. }\label{model}
\end{figure*}

Recently, the topolectrical circuit springs up as a powerful platform to study the fundamental topological physics \cite{Lee2018,Imhof2018,Hofmann2019,Zhu2019,Lu2019,Yyt2020,Yang2020,Song2020,Ezawa2020,Ezawa20202}, since simple inductor-capacitor (LC) networks can fully simulate the tight-binding model in condensed matter physics. In this Letter, we fabricate two kinds of Kekul\'{e} LC circuits with molecule-zigzag and partially-bearded edges (see Fig. \ref{model}). By measuring the node-ground impedance and monitoring the spatiotemporal voltage signal propagation, we observe the quantum pseudospin Hall effect emerging in the opposite parameter regions with flipped helicities for the two different edge terminations, where the chirality of the circulating current in the unit cell mimics the spin. Quantized mirror winding number is proposed to explain our experimental findings.

We consider two finite-size artificial Kekul\'{e} circuits with molecule-zigzag and partially-bearded edge terminations, as shown in Figs. \ref{model} (a) and \ref{model}(b), respectively. The circuits consist of two types of capacitors $C_A$, $C_B$ and inductor $L$. The response of the circuit at frequency $\omega$ is given by Kirchhoff's law:
\begin{equation} \label{KL}
I_a(\omega)=\sum_bJ_{ab}(\omega)V_b(\omega),
\end{equation}
where $I_a$ is the external current flowing into node $a$, $V_b$ is the voltage of node $b$, and $J_{ab}(\omega)=i\omega\left[C_{ab}+\delta_{ab}(\sum_nC_{an}-\frac{1}{\omega^2L_a})\right]$ is the circuit Laplacian, with $C_{ab}$ the capacitance between nodes $a$ and $b$. Based on Eq. \eqref{KL}, one can explicitly express the circuit Laplacian $J_{\rm I}(\omega)$ and $J_{\rm II}(\omega)$ of the two circuits in Figs. \ref{model}(a) and \ref{model}(b) \cite{SM}. At the resonant frequency $\omega_0=1/\sqrt{(2C_A+C_B)L}$, the diagonal elements of circuit Laplacians vanish, and the circuit model is equivalent to the tight-binding model with $-\omega_0C_A$ and $-\omega_0C_B$ being two hopping coefficients.

\begin{figure*}
  \centering
  \includegraphics[width=0.9\textwidth]{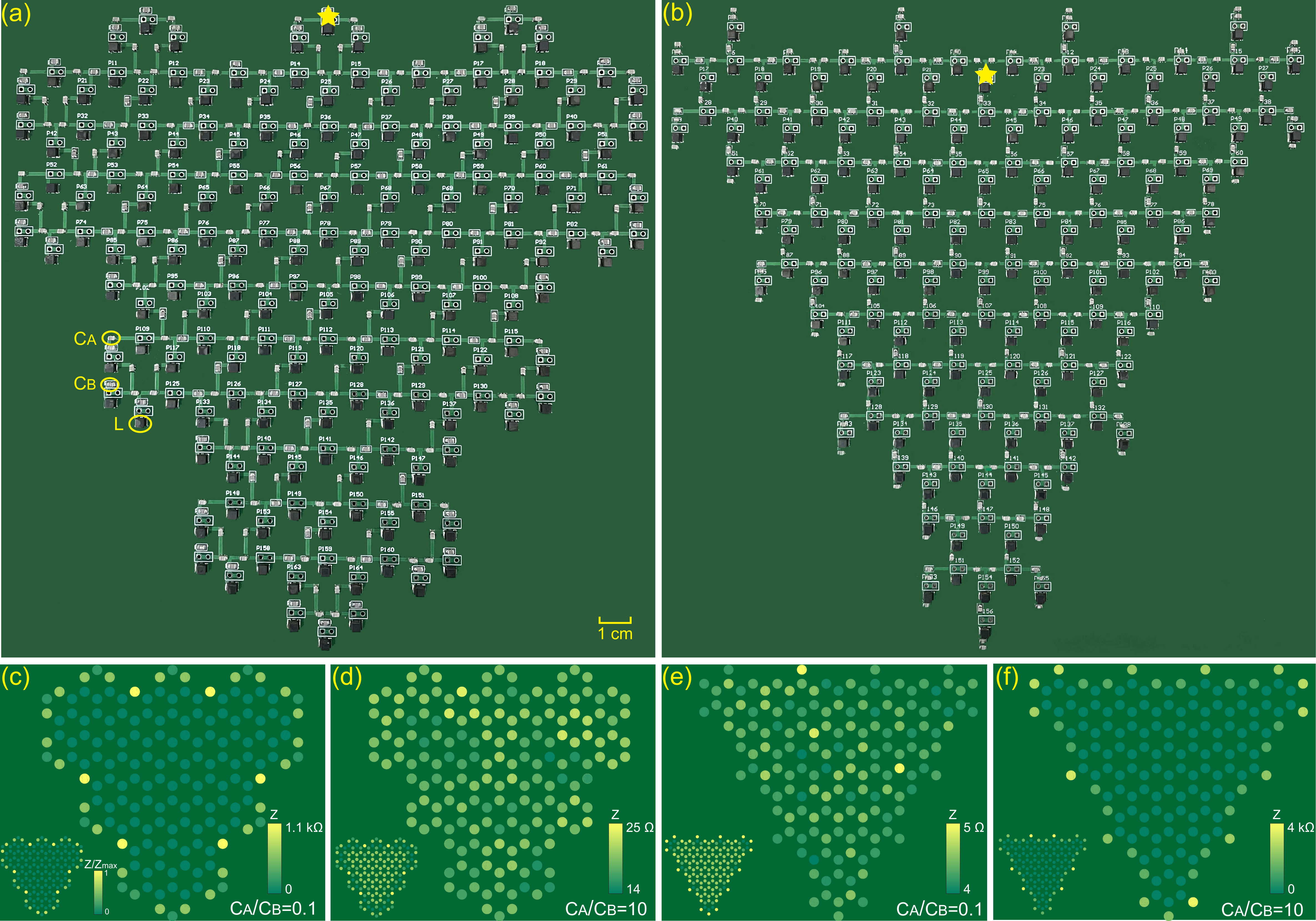}\\
  \caption{Printed circuit boards with (a) molecule-zigzag and (b) partially-bearded edges. Yellow stars indicate the position of signal sources in the voltage measurements. (c)-(f) Experimental measurements of the spatial distribution of impedance between each node and the ground. Insets: numerical results.}\label{Z}
\end{figure*}

We fabricate two printed circuit boards with different edge geometries displayed in Figs. \ref{Z}(a) and \ref{Z}(b), respectively. In experiments, we adopt $C_A = 1$ nF, $C_B =10$ nF or $0.1$ nF, and $L=39~\mu$H (all circuit elements have a $2\%$ tolerance), with the resonant frequency being $\omega_{0}/2\pi=1/[2\pi\sqrt{(2C_A+C_B)L}]=232.65$ kHz or $556.13$ kHz, respectively.

We measure the distributions of impedance between each node and the ground by an analyser (Keysight E4990A), with the results plotted in Figs. \ref{Z}(c)-\ref{Z}(f). For devices with molecule-zigzag edge at $C_A/C_B=0.1$ [Fig. \ref{Z}(c)] and partially-bearded edge at $C_A/C_B=10$ [Fig. \ref{Z}(f)], we observe that the impedance concentrates on the sample edge, the value of which is larger than one thousand Ohms, indicating the existence of edge states. Theoretically, the impedance between node $a$ and $b$ is given by \cite{Yang2020}:
\begin{equation}\label{Zab}
Z_{ab}=\frac{V_a-V_b}{I_{ab}}=\sum_{n}\frac{|\psi_{n,a}-\psi_{n,b}|^2}{j_n},
\end{equation}
where $|\psi_{n,a}-\psi_{n,b}|$ is the amplitude difference between $a$ and $b$
nodes of the $n$th eigenstate, and $j_n$ is the $n$-th eigenvalue. We plot the  numerical results in the insets of Figs. \ref{Z}(c)-\ref{Z}(f), showing an excellent agreement with the experimental measurements.

\begin{figure*}
  \centering
  \includegraphics[width=0.9\textwidth]{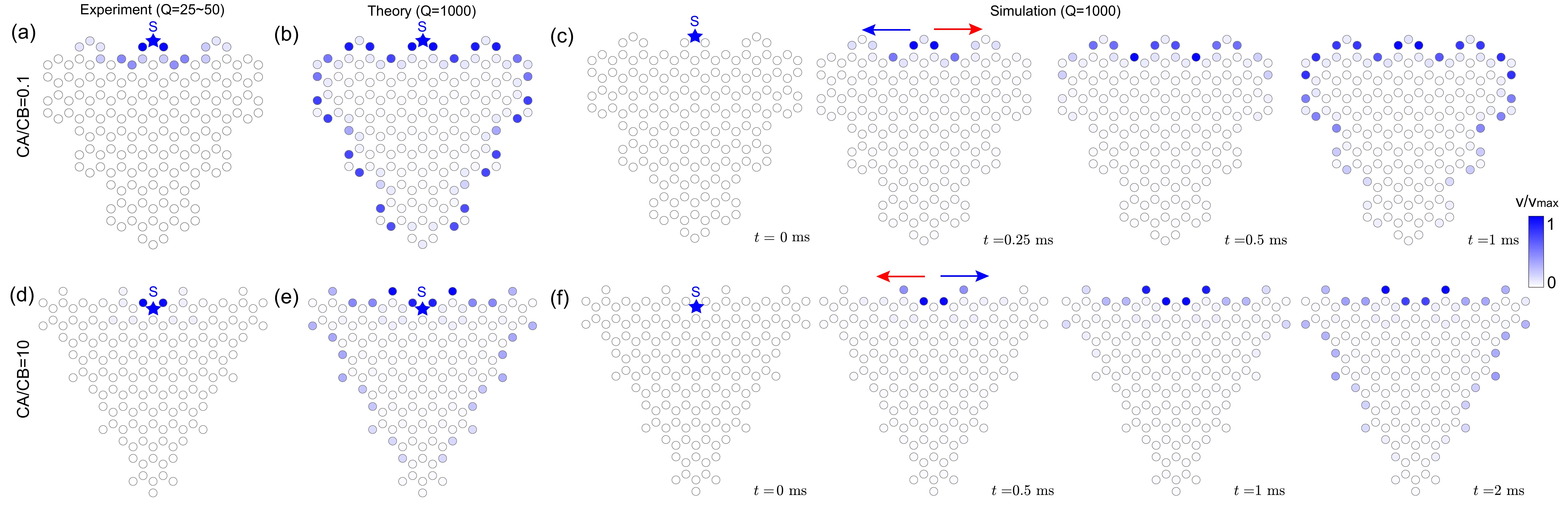}\\
  \caption{Experimental measurements of the steady-state voltage distribution in the devices with (a) molecule-zigzag ($C_A/C_B=0.1$) and (d) partially-bearded ($C_A/C_B=10$) edges. (b)(e) Theoretical calculation with a higher $Q$-factor ($Q=1000$).  (c)(f) Snapshots of the propagating voltage signal at different times, with the blue star indicating the position of the signal source, and the red and blue arrows representing the propagation direction of the voltage signal with pseudospin up and down, respectively.}\label{WF}
\end{figure*}

It's known that the QSH insulator allows bidirectional propagation states along the boundary. However, we cannot directly observe the time-resolved wave dynamics by measuring the impedance. To solve this problem, we monitor and record the spatiotemporal voltage signal in the circuits. Specifically, we impose a sinusoidal voltage signal $v(t)=v_0\sin(\omega_0 t)$ with the amplitude $v_0=5$ V at the node labeled by blue stars in Figs. \ref{WF}(a) and \ref{WF}(b) by an arbitrary function generator (GW AFG-3022), and then measure the steady-state voltage distribution by the oscilloscope (Keysight MSOX3024A). We indeed observe a strong voltage response along both directions of the device edge. It is noted that the voltage signal decays very fast away from the voltage source, because of the low quality factor ($Q=25-50$) of the inductors. In Figs. \ref{WF}(b) and \ref{WF}(e), we plot the theoretical steady-state voltage distributions with higher $Q$-factor inductors (we set $Q=1000$, realized by introducing a small resistance to each inductor), which improves the visualization of the bidirectional edge states.

To see the propagation details of the edge states, we perform circuit simulations with LTspice \cite{LTspice} and record the voltage of all nodes. For the two edge states along molecule-zigzag and partially-bearded boundaries, the voltage signals propagate in both directions along the edge, as displayed in Figs. \ref{WF}(c) and \ref{WF}(f), accompanied by a helicity flipping indicated by red and blue arrows (see analysis below).

\begin{figure*}
  \centering
  \includegraphics[width=0.9\textwidth]{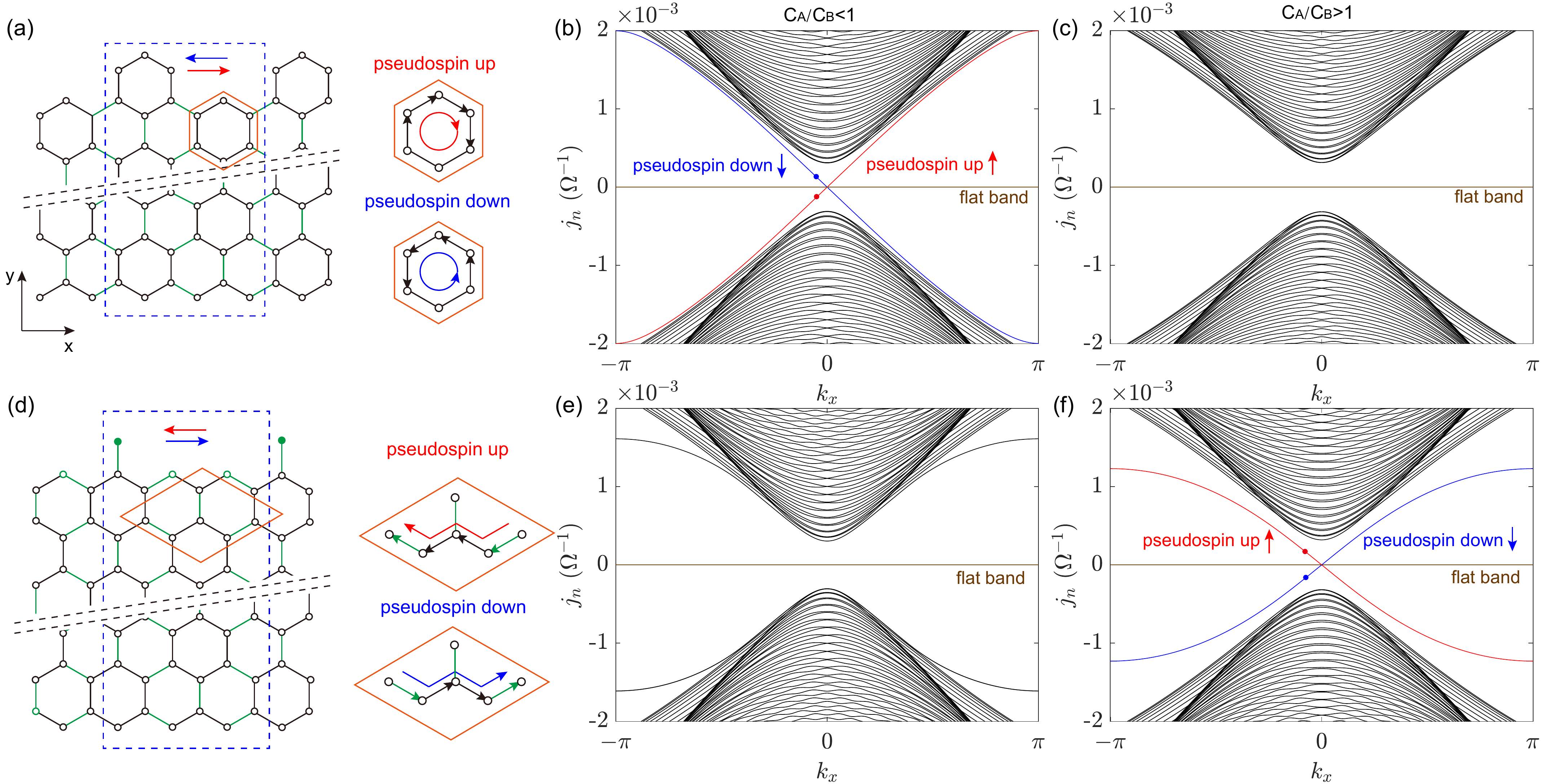}\\
  \caption{(a) Schematic plot of a ribbon with molecule-zigzag edge (top) and graphene-zigzag edge (bottom). The ribbon is periodic along $\hat{x}$ direction and contains 40 unit cells along $\hat{y}$ direction. Insets: the pesudospin is denoted by the chirality of the circulating current in the unit cell. The band structure of the ribbon with two different capacitor ratios: (b) $C_A:C_B=1:1.1$ and (c) $C_A:C_B=1:0.9$. Red and blue lines represent the dispersive edge states with pesudospin up and down counterpropagating along the top edge. Brown line denotes the localized edge mode in the bottom boundary. (d) Illustration of a ribbon with partially-bearded edge (top) and graphene-zigzag edge (bottom). The band structure of the ribbon with two different capacitor ratios: (e) $C_A:C_B=1:1.1$ and (f) $C_A:C_B=1:0.9$.}\label{ribbon}
\end{figure*}

To explain the experimental results, we numerically calculate the band structure of the circuits. By diagonalizing the circuit Laplacians $J_{\rm I}(\omega)$ and $J_{\rm II}(\omega)$, we obtain the admittance spectrum $j_n$ and the corresponding wave functions $\psi_{n,m}$, shown in Fig. S1 in Supplemental Material \cite{SM}. For circuits of molecule-zigzag edge, with $C_A/C_B=0.1$, isolated states emerge in the gap of the bulk admittance spectrum, which correspond to the edge states. When $C_A/C_B=10$, only are bulk states identified. For circuits of partially-bearded edge, on the contrary, we find that the edge states emerge in the opposite capacitance ratio, i.e., $C_A/C_B=10$. For $C_A/C_B=0.1$, one can only observe the bulk states. These results are fully consistent with our experimental observations.

Next, we analyze the origin of the bidirectional edge states. First of all, we can exclude the Tamm-Shockley mechanism \cite{Tamm1932,Shockley1939}, which predicts that the periodicity breaking of the crystal potential at the boundary can lead to the formation of a conducting surface/edge state. However, this surface/edge state is trivial because it is sensitive to impurities, defects, and disorder, which is not compatible with our experimental findings. There thus must be a topological reason for the emerging bidirectional edge states we observed. To justify this point of view, we employ the mirror winding number $(n_+,n_-)$ defined in the unit cell with
\begin{equation} \label{winding_number}
n_\pm=-\frac{1}{2\pi}\oint\frac{d}{dk_\perp}\arg(\det Q_{k^\pm_\perp})d k_\perp
\end{equation}
in the presence of chiral symmetry. The analytical expression of matrices $Q_{k_\perp^{\pm}}$ can be found in Sec. II of Supplemental Material \cite{SM}. The choice of the unit cell depends on the shape of sample edge. As shown in Figs. \ref{model}(a) and \ref{model}(b), the dashed red hexagon and rhombus represent the approximate unit cells for the two different edge geometries, respectively. For the circuit with molecule-zigzag edge, we obtain $(n_+,n_-)=(1,-1)$ when $C_A/C_B<1$ and $(0,0)$ when $C_A/C_B>1$. Therefore, we can observe the topological edge states when $C_A/C_B<1$. For the circuit with partially-bearded edge, the case is adverse to the former: $(n_+,n_-)=(0,0)$ when $C_A/C_B<1$ and $(1,-1)$ when $C_A/C_B>1$, indicating that the topological edge states arise in the region of $C_A/C_B>1$ \cite{SM}.

Figures \ref{ribbon}(a) and \ref{ribbon}(d) show two infinite-long ribbons with molecule-zigzag and partially-bearded edges. For the ribbon with molecule-zigzag edge, in the case of $C_A/C_B<1$, we find three isolated modes in the band gap [see Fig. \ref{ribbon}(b)]. The red and blue spectrums represent the helical edge states because of the opposite group velocity. Interestingly, we can define the circulating bond currents inside the unit cell: $i_{m\rightarrow n}={\rm Im} [\psi_m^*\psi_n]$ \cite{Zhang2008,Zhang2009,Wu2016} with their flowing direction plotted in the right side of Figs. \ref{ribbon}(a) and \ref{ribbon}(d). We find that the chirality of the circulating current in the unit cell are opposite for the in-gap red and blue bands, which mimics the electron spin-up and spin-down states, respectively. This observation is reminiscent of the spin-momentum locking in the QSH effect. Brown line denotes the flat band localized in the bottom zigzag edge of the ribbon \cite{Fujita1996}. For $C_A/C_B>1$, there is no in-gap energy spectrum expect for the flat band, see Fig. \ref{ribbon}(c). For the ribbon with partially-bearded edge, the edge modes with flipped helicity however only appear in the region of $C_A/C_B>1$ [see Figs. \ref{ribbon}(e) and \ref{ribbon}(f)]. These results well explain the numerical calculations and experimental measurements.

To understand the helicity flipping, we map the six-band circuit model to the four-band Bernevig-Hughes-Zhang (BHZ) model originally proposed for HgTe quantum wells \cite{Bernevig2006,Konig2007}. To this end, we express $J_{ab}(\omega)=i\mathcal{H}_{ab}(\omega)$, in which $\mathcal{H}(\omega)$ can be viewed as a hermitian tight-binding Hamiltonian. Taking the molecule-zigzag unit cell as an example, one can write the Hamiltonian of an infinite Kekul\'{e} circuit at resonant frequency as below:
\begin{equation} \label{H1}
\mathcal{H}=-\omega_0 C_A\sum_{\left<i,j\right>}c_i^\dagger c_j-\omega_0 C_B\sum_{\left<i',j'\right>}c_{i'}^\dagger c_{j'},
\end{equation}
where $c_i$ is the annihilation operator at site $i$, and $\left<i,j\right>$ and $\left<i',j'\right>$ run over nearest-neighboring sites inside and between hexagonal unit cells, respectively. Diagonalizing Hamiltonian \eqref{H1}, we obtain six bands, two of which are high-energy bands with the phase transition point $C_A/C_B=1$ at the low-energy $\Gamma$ point, as shown in Fig. S3 in Supplemental Material \cite{SM}. We further note that the high-energy parts are irrelevant to the topological phase transition. By performing a unitary transformation $\mathcal{H'}=U^\dagger \mathcal{H} U$ on $\mathcal{H}$ around the $\Gamma$ point \cite{SM}, we separate the two high-energy orbits and obtain the low-energy effective BHZ-type Hamiltonian as:
\begin{equation}
\mathcal{H}_{\rm eff}({\bf k})=-\omega_0\left(
                        \begin{array}{cc}
                            H(k) & 0 \\
                            0 & H^*(-k) \\
                        \end{array}
                      \right),\\
\end{equation}
with $H(k)=\left(
                        \begin{array}{cc}
                        M-Bk^2 & Ak_- \\
                        A^*k_+ & -M+Bk^2 \\
                           \end{array}
                       \right),$ where $M=C_B-C_A$, $A=-\frac{3}{2}iC_B$, $B=\frac{9}{4}C_B$, $k^2=k_x^2+k_y^2$, and $k_{\pm}=k_x{\pm}ik_y$.

For the circuit with partially-bearded unit cell, we get the similar low-energy effective Hamiltonian, but with a different $M=C_A-C_B$. The sign of parameter $M$ is opposite for the two edge geometries, leading to the helicity flipping of the edge states in the opposite parameter regions based on the band inversion mechanism. We thus conclude that, although Kirchhoff's law is rather different from the Sch\"{o}rdinger equation, the underlying physics between our circuit model and the quantum well model is actually quite similar. The parameter $M$ can be viewed as an effective spin-orbit coupling (SOC) associated with the pseudo spin, which is different from the intrinsic one originating from the relativistic effect. Whereas, the SOC in circuit is more controllable and can be very large, enabling the observation of the quantum pseduo-spin Hall states at room temperature.

In summary, we reported an edge-dependent quantum pseudospin Hall effect in topolectric circuits. We showed that the pesudospin is represented by the chirality of the circulating current in the unit cell. Through the impedance measurement and spatiotemporal voltage signal detection assisted by circuit simulations, we directly identified the helical nature of the edge states. The emerging topological phases were characterized by mirror winding numbers, which depend on the shape of device edge. Our work uncovers the importance of the edge geometry on the QSH effect, and opens a new pathway of using circuits to simulate the spin-dependent topological physics, that may inspire research in other solid-state systems in the future.

\begin{acknowledgments}
This work was supported by the National Natural Science Foundation of China (Grants No. 12074057, No. 11604041, and No. 11704060). X. R. Wang acknowledges the financial support of Hong Kong RGC (Grants No. 16300117, 16301518, and 16301619).
\end{acknowledgments}

\newpage
\begin{widetext}

\begin{flushleft}
\center{\LARGE{\textbf{Supplemental Material}}}
\\[0.5cm]

{\Large{\textbf{Experimental observation of edge-dependent quantum pseudospin Hall effect}}}
\quad\par
\quad\par
Huanhuan Yang$^1$, Lingling Song$^1$, Yunshan Cao$^1$, X. R. Wang$^{2}$,$^{*}$ and Peng Yan$^{1\dagger}$
\quad\par
\quad\par

$^{1}$\emph{\small{School of Electronic Science and Engineering and State Key Laboratory of Electronic Thin Films and Integrated Devices, University of Electronic Science and Technology of China, Chengdu 610054, China and}}

$^{2}$\emph{\small{Physics Department, The Hong Kong University of Science and Technology, Clear Water Bay, Kowloon, Hong Kong}}

\end{flushleft}

\section{I. Circuit Laplacian}
In this section, we show the circuit Laplacian of the two circuits in the main text. For the circuit with molecule-zigzag edge geometry:
\begin{equation}
J_{\rm I}(\omega)=\omega \left(
\begin{array}{ccccccc}
J_0  & 0  & 0 &  -J_A   &   -J_A   &0 &\ldots\\
0 & J_0   & 0 & 0 &   0   & -J_A&\ldots \\
-J_A & 0  & J_0  &  0   &  0 & 0&\ldots\\
-J_A &  0 & 0    &  J_0 &  0    & 0&\ldots\\
0    & 0     & 0 &  0   &  J_0  & 0&\ldots\\
0    & -J_A     & 0 &  0   &   0 & J_0&\ldots\\
\vdots  & \vdots  & \vdots &  \vdots   &  \vdots  & \vdots&\ddots \\
\end{array}
\right)_{168\times 168},
\end{equation}
with $J_0=2C_A+C_B-1/(\omega^2L)$, $J_A=C_A$, and $J_B=C_B$.
For the circuit with partially-bearded edge geometry:
\begin{equation}
J_{\rm II}(\omega)=\omega \left(
\begin{array}{ccccccc}
J_0  & 0  & 0 &  0   &   0   &-J_A &\ldots\\
0 & J_0   & 0 & 0 &   0   & 0&\ldots \\
0 & 0  & J_0  &  0   &  0 & 0&\ldots\\
0 &  0 & 0    &  J_0 &  0    & 0&\ldots\\
0    & 0     & 0 &  0   &  J_0  & 0&\ldots\\
-J_A    & 0     & 0 &  0   &  0  & J_0&\ldots\\
\vdots  & \vdots  & \vdots &  \vdots   &  \vdots  & \vdots&\ddots \\
\end{array}
\right)_{156\times 156}.
\end{equation}

Diagonalizing $J_{\rm I}(\omega)$ and $J_{\rm II}(\omega)$, we obtain the admittance spectrum $j_n$ and the corresponding wave functions $\psi_{n,m}$. To directly compare with the experimental results, we adopt $C_A=1$ nF, $L=39$ $\mu$H, and $C_B =10$ nF or $0.1$ nF. The admittance spectrums with the insets show the typical profiles of wave functions are displayed in Fig. \ref{EIG}(a)-\ref{EIG}(d). For the circuits with molecule-zigzag edges, in the case of $C_A/C_B=0.1$, we find a series of isolated states in the gap of the admittance spectrum (blue dots), which correspond to the helical edge states, shown in Fig. \ref{EIG}(a). We confirm that all blue dots are edge states (not shown). In the regime of $C_A/C_B=10$, only the bulk states exist, see Fig. \ref{EIG}(b). However, for the circuits with partially-bearded edges, we find that the edge states emerge in the opposite region, i.e., $C_A/C_B=10$, as shown in Fig. \ref{EIG}(d). In the case of $C_A/C_B=0.1$, we can only see the bulk states, see Fig. \ref{EIG}(c).

\begin{figure}[htbp!]
  \centering
  \includegraphics[width=1\textwidth]{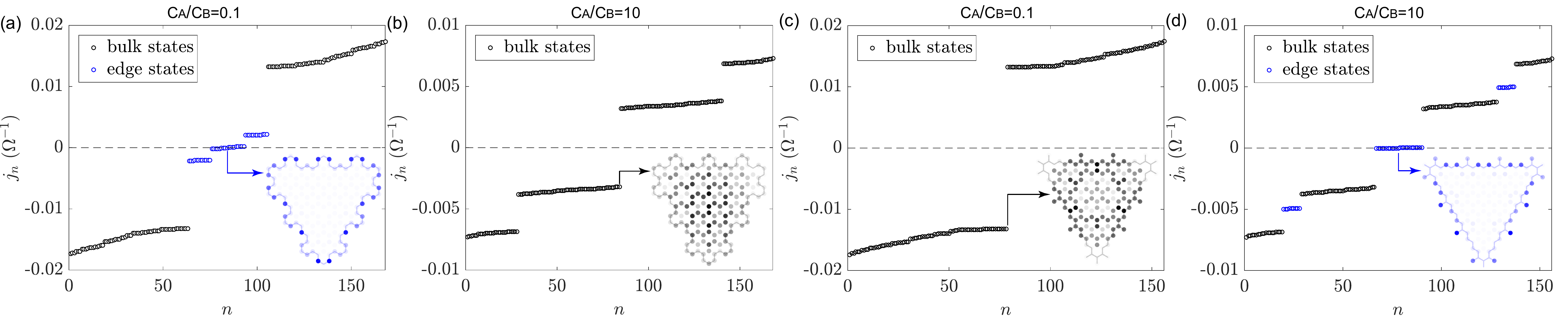}\\
  \caption{Admittance spectrum at different edges and parameters. The blue and black dots denote the edge states and bulk states, respectively. Insets: spatial distribution of wave functions with the number of state indicated by the arrows.  (a)(b) molecule-zigzag edge with $C_A/C_B=0.1$ and $C_A/C_B=10$. (c)(d) partially-bearded edge with $C_A/C_B=0.1$ and $C_A/C_B=10$.}\label{EIG}
\end{figure}

\section{II. Mirror winding number}
In this section, we calculate the topological invariant mirror winding number to characterize the helical edge states. If we express $J_{ab}(\omega)=i\mathcal{H}_{ab}(\omega)$, $\mathcal{H}(\omega)$ can be viewed as a tight-binding Hamiltonian. With the appropriate unit cells in Fig. \ref{unit_cell} (unit cell I for the circuit with molecule-zigzag edge, and unit cell II for the circuit with partially-bearded edge), one can write the Hamiltonian of an infinite Kekul\'{e} circuit as:
\begin{equation} \label{H}
\mathcal{H}=\omega\left(
              \begin{array}{cccccc}
                h_{0} & 0 & 0 &  &  &  \\
                0 & h_{0} & 0 &  & -Q_{\bf k} &  \\
                0 & 0 & h_{0} &  &  &  \\
                 &  &  & h_{0} & 0 & 0 \\
                 & -Q_{\bf k}^\dagger &  & 0 & h_{0} & 0 \\
                 &  &  & 0 & 0 & h_{0} \\
              \end{array}
            \right),
\end{equation}
with the matrix elements $h_0=2C_A+C_B-1/(\omega^2L)$,
\begin{equation} \label{Qk1}
Q^{\rm I}_{\bf k}=\left(
            \begin{array}{ccc}
              C_BX\overline{Y}^2 & C_A & C_A \\
              C_A & C_B\overline{X}Y & C_A \\
              C_A & C_A & C_BY \\
            \end{array}
          \right)
\end{equation}
for molecule-zigzag edge, where $X=e^{i{\bf k}\cdot{\bf a}_1}$, $Y=e^{i{\bf k}\cdot{\bf a}_2}$ with ${\bf a}_1=3\sqrt{3}\hat{x}$ and ${\bf a}_2=\frac{3\sqrt{3}}{2}\hat{x}+\frac{3}{2}\hat{y}$ being the two basic vectors, and
\begin{equation} \label{Qk2}
Q^{\rm II}_{\bf k}=\left(
            \begin{array}{ccc}
              C_B & C_A & C_A \\
              C_A\overline{Y} & C_B & C_A\overline{X}Y \\
              C_AX\overline{Y} & C_AY & C_B \\
            \end{array}
          \right)
\end{equation}
for partially-bearded edge.

\begin{figure}[htbp!]
  \centering
  \includegraphics[width=0.7\textwidth]{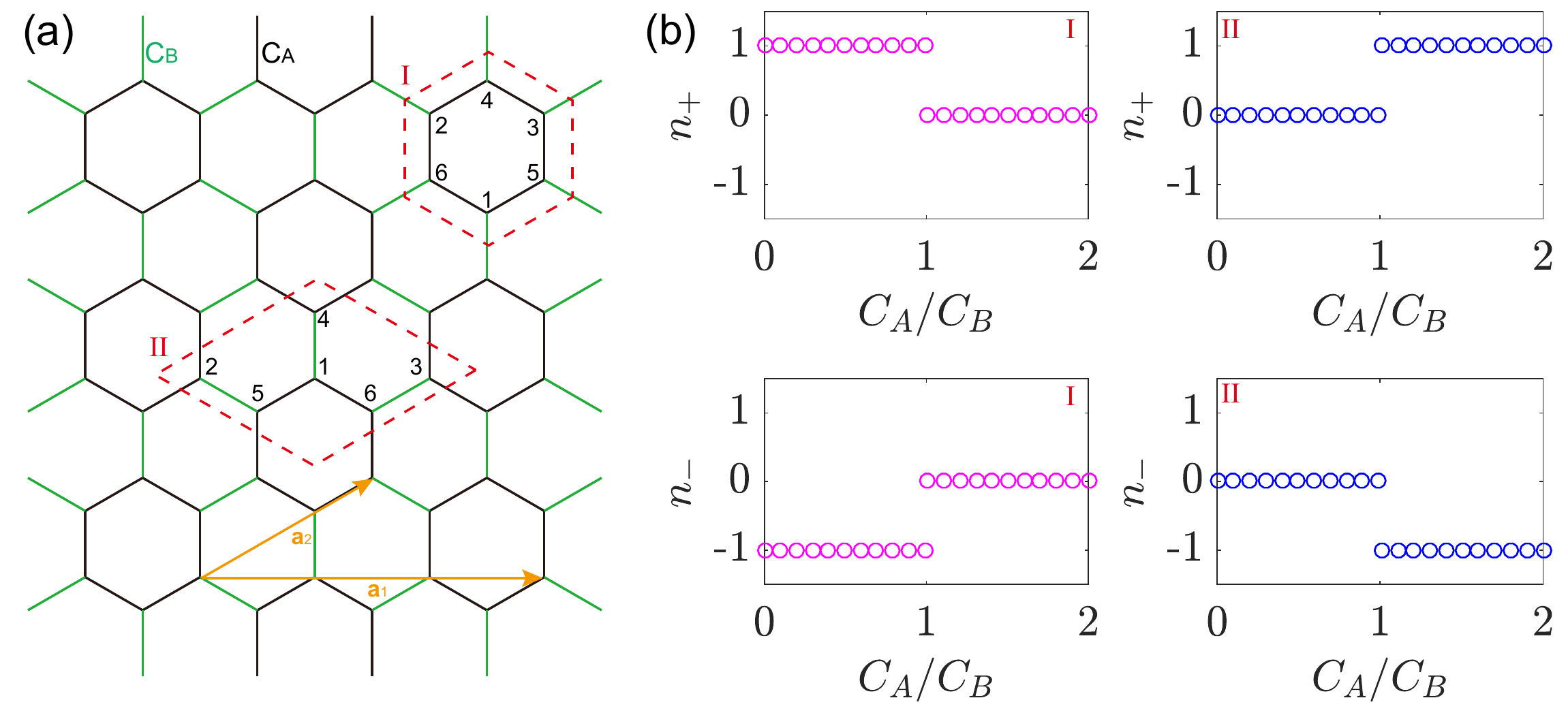}\\
  \caption{(a) Appropriate unit cells for molecule-zigzag and partially-bearded edges. The orange arrows indicate the two basic vectors. (b) The mirror winding numbers $(n_+,n_-)$ as a function of the capacitance ratio $C_A/C_B$. }\label{unit_cell}
\end{figure}

At resonant frequency $\omega_0=1/\sqrt{(2C_A+C_B)L}$, the diagonal element $h_0$ vanishes, and the  Hamiltonian can be simplified as:
\begin{equation} \label{Hs}
\mathcal{H}=-\omega_0\left(
              \begin{array}{cc}
                0 & Q_{\bf k} \\
                Q^\dagger_{\bf k} & 0 \\
              \end{array}
            \right),
\end{equation}
where $Q_{\bf k}$ is $Q^{\rm I}_{\bf k}$ (Eq. \ref{Qk1}) for molecule-zigzag edge, and $Q^{\rm II}_{\bf k}$ (Eq. \ref{Qk2}) for partially-bearded edge.

Regarding the momentum $\bf k$ parallel to the unit vector ${\bf a}_1$ defined as a free parameter, the system can be viewed as an effective 1D model, to which one can assign the winding number as:
\begin{equation} \label{winding_number}
n(k_\parallel)=-\frac{1}{2\pi}\oint\frac{d}{dk_\perp}\arg(\det Q_{k_\parallel,k_\perp})d k_\perp
\end{equation}

For $k_\parallel=0$, the mirror symmetry with the mirror plane perpendicular to ${\bf a}_1$ enables us to decompose the Hamiltonian \eqref{Hs} into even and odd sectors $H_{k_\perp^{\pm}}$, where {\bf k} is replaced by $k_\perp$. Concretely, $Q_{\bf k}$ can be decomposed into even and odd sectors $Q_{k_\perp^{\pm}}$. Then, we can assign winding numbers for the even and odd sectors separately by substituting $Q_{k_\perp^{+}}$ and $Q_{k_\perp^{-}}$ into Eq. \ref{winding_number}, which constitutes the mirror winding number $(n_+, n_-)$  \cite{Kariyado2017}.

At $k_\parallel = 0$, $Q^{\rm I}_{\bf k}$ is decomposed into
\begin{equation}
Q^{\rm I}_{k_\perp^+}=\left(
                        \begin{array}{cc}
                          C_B\overline{Y}^2 & \sqrt{2}C_A \\
                          \sqrt{2}C_A & C_A+C_BY \\
                        \end{array}
                      \right),~Q^{\rm I}_{k_\perp^-}=C_BY-C_A,
\end{equation}
and  $Q^{\rm II}_{\bf k}$ is decomposed into
\begin{equation}
Q^{\rm II}_{k_\perp^+}=\left(
                        \begin{array}{cc}
                          C_B & \sqrt{2}C_A \\
                          \sqrt{2}C_A\overline{Y} & C_B+C_AY \\
                        \end{array}
                      \right),~Q^{\rm II}_{k_\perp^-}=C_B-C_AY.
\end{equation}

Using Eq. \ref{winding_number}, we can compute the mirror winding number $(n_+, n_-)$ immediately, with the results plotted in Fig. \ref{unit_cell}(b). For the circuit with molecule-zigzag and partially-bearded edge, the topological edge states appear in the region of $C_A/C_B<1$ and $C_A/C_B>1$ respectively.

\section{III. Analogy to the Quantum Spin Hall Effect}

In this section, we map our six-band circuit model to the four-band Bernevig-Hughes-Zhang (BHZ) model for CdTe/HgTe/CdTe quantum wells.
\begin{figure}[htbp!]
  \centering
  \includegraphics[width=0.9\textwidth]{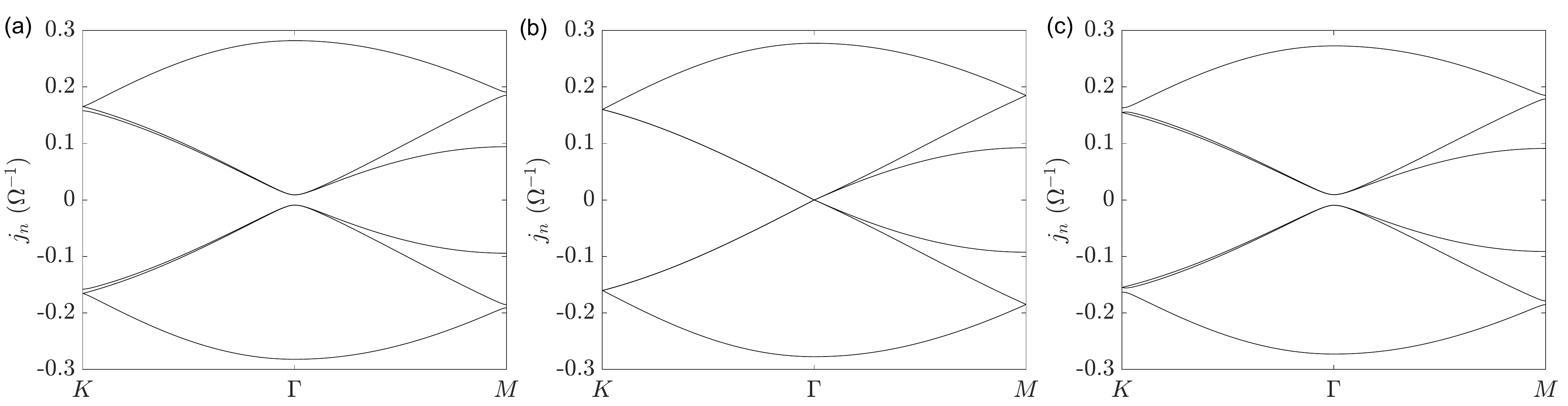}\\
  \caption{Admittance spectrum for different capacitance ratio. (a) $C_A/C_B=0.9$, (b) $C_A/C_B=1$, and (c) $C_A/C_B=1.1$.}\label{ES}
\end{figure}

We calculate the energy spectrum of Eq. \ref{H} for three capacitance ratios, plotted in Fig. \ref{ES}. The spectrum are gapless when $C_A/C_B = 1$, and the phase transition point is at the $\Gamma$ point.  Two bands of the spectrum are high-energy parts, which are irrevelant to the topological phase transition. Therefore, the six-band Hamiltonian \eqref{H} can be downfolded into the four-band one by omitting the two high-energy bands \cite{Yang2020}.

Taking the circuit with molecule-zigzag edge geometry as an example, we impose a unitary transformation $\mathcal{H'}=U^\dagger \mathcal{H} U$ on Hamiltonian $\mathcal{H}$ \eqref{H} to separate the high-energy parts of the Hamiltonian with the matrix:
\begin{equation}
\mathcal{U}=\frac{1}{\sqrt{6}}\left(
              \begin{array}{cccccc}
    e^{i\frac{\pi}{2}}&e^{i\pi}&e^{i\frac{3\pi}{2}}&e^{i\pi}&1&1\\
    e^{i\frac{7\pi}{6}} &e^{i\frac{\pi}{3}} &e^{i\frac{5\pi}{6}}& e^{i\frac{5\pi}{3}}& -1& 1\\
    e^{i\frac{11\pi}{6}} &e^{i\frac{5\pi}{3}}& e^{i\frac{\pi}{6}}& e^{i\frac{\pi}{3}}& 1 &1\\
    e^{i\frac{\pi}{2}}& e^{i2\pi}& e^{i\frac{3\pi}{2}} &e^{i2\pi}& -1 &1\\
    e^{i\frac{7\pi}{6}}& e^{i\frac{4\pi}{3}}& e^{i\frac{5\pi}{6}}& e^{i\frac{2\pi}{3}}& 1 &1\\
    e^{i\frac{11\pi}{6}}& e^{i\frac{2\pi}{3}}& e^{i\frac{\pi}{6}}& e^{i\frac{4\pi}{3}}& -1 &1\\
              \end{array}
            \right).
\end{equation}
Then, imposing Taylor expansion on each matrix element of $\mathcal{H}'$ around the $\Gamma$ point to 2nd-order terms, we obtain:
\begin{equation} \label{HG}
\mathcal{H}_{\Gamma}=-\omega_0\left(
  \begin{array}{cccccc}
    \delta C-\frac{9}{4}C_Bk^2 & -\frac{3}{2}iC_Bk_- & h_{13} & 0 &-\frac{3}{2}iC_Bk_+ & h_{16}\\
    \frac{3}{2}iC_Bk_+ & -\delta C+\frac{9}{4}C_Bk^2 & 0 & h_{24}  & h_{25} & -\frac{3}{2}C_Bk_- \\
    h_{13}^* & 0 & \delta C-\frac{9}{4}C_Bk^2 & -\frac{3}{2}iC_Bk_+ & -\frac{3}{2}iC_Bk_- & h_{36} \\
    0 & h_{24}^* & \frac{3}{2}iC_Bk_-& \delta C+\frac{9}{4}C_Bk^2 & h_{45} & \frac{3}{2}C_Bk_+ \\
    \frac{3}{2}iC_Bk_- & h_{25}^* & \frac{3}{2}iC_Bk_+ & h_{45}^* & -2C_A-C_B+\frac{9}{4}C_Bk^2 & 0 \\
    h_{16}^* & -\frac{3}{2}C_Bk_+ & h_{36}^* & \frac{3}{2}C_Bk_- & 0 & 2C_A+C_B-\frac{9}{4}C_Bk^2 \\
  \end{array}
\right)
\end{equation}
with $\delta C=C_B-C_A$, $k^2=k_x^2+k_y^2$, $k_-=k_x-ik_y$, $h_{13}=h_{24}^*=\frac{9}{8}C_B(k_y^2-k_x^2)-\frac{9}{4}C_Bik_xk_y$, $h_{16}= \frac{9}{8}C_Bi(k_y^2-k_x^2)-\frac{9}{4}C_Bk_xk_y$, $h_{25}=h_{45}^*=\frac{9}{8}C_B(k_x^2-k_y^2)+\frac{9}{4}C_Bik_xk_y$, and $h_{36}=\frac{9}{8}C_Bi(k_x^2-k_y^2)-\frac{9}{4}C_Bk_xk_y$.

Dropping the last two high-energy orbits and the second-order off-diagonal terms $h_{ij}$ ($h_{ij}$ contribute as high-order perturbations), Hamiltonian \eqref{HG} is block diagonalized. We obtain the low-energy effective Hamiltonian as:
\begin{equation}
\mathcal{H}^{\rm I}_{\rm eff}=-\omega_0\left(
  \begin{array}{cccc}
    \delta C-\frac{9}{4}C_Bk^2 & -\frac{3}{2}iC_Bk_- & 0 & 0 \\
    \frac{3}{2}iC_Bk_+ & -\delta C+\frac{9}{4}C_Bk^2 & 0 & 0   \\
    0 & 0 & \delta C-\frac{9}{4}C_Bk^2 & -\frac{3}{2}iC_Bk_+  \\
    0 & 0 & \frac{3}{2}iC_Bk_-& \delta C+\frac{9}{4}C_Bk^2  \\
  \end{array}
\right).
\end{equation}

The effective Hamiltonian $\mathcal{H}^{\rm I}_{\rm eff}$ can be rewritten in a concise BHZ form \cite{Bernevig2006} as:
\begin{equation}  \label{BHZ}
\begin{aligned}
\mathcal{H}_{\rm eff}(k)=-\omega_0\left(
                        \begin{array}{cc}
                            H(k) & 0 \\
                            0 & H^*(-k) \\
                        \end{array}
                      \right),~
{\rm with}~H(k)&=\left(
                        \begin{array}{cc}
                        M-Bk^2 & Ak_- \\
                        A^*k_+ & -M+Bk^2 \\
                           \end{array}
                       \right),
\end{aligned}
\end{equation}
where $M=\delta C=C_B-C_A$, $A=-\frac{3}{2}iC_B$, $B=\frac{9}{4}C_B$, $k^2=k_x^2+k_y^2$, and $k_{\pm}=k_x{\pm}ik_y$.

Similarly, near the $\Gamma$ point, the Hamiltonian \eqref{Hs} with the partially-bearded edge geometry can be simplified as:
\begin{equation}
\mathcal{H}^{\rm II}_{\rm eff}=-\omega_0\left(
                        \begin{array}{cccc}
                           \delta C+\frac{9}{4}C_Ak_x^2+\frac{3}{4}C_Ak_y^2 & \frac{3}{2}iC_Ak_- & 0 & 0 \\
                          -\frac{3}{2}iC_Ak_+ &  -\delta C-\frac{9}{4}C_Ak_x^2-\frac{3}{4}C_Ak_y^2 & 0 & 0 \\
                          0 & 0 &  \delta C+\frac{9}{4}C_Ak_x^2+\frac{3}{4}C_Ak_y^2 & \frac{3}{2}iC_Ak_+\\
                          0 & 0 & -\frac{3}{2}iC_Ak_- &  -\delta C-\frac{9}{4}C_Ak_x^2-\frac{3}{4}C_Ak_y^2 \\
                        \end{array}
                      \right),
\end{equation}
which can be rewritten as the BHZ form with
\begin{equation}
H(k)=-\left(\begin{array}{cc}
       M-B(3k_x^2+k_y^2) & Ak_- \\
            A^*k_+ & -M+B(3k_x^2+k_y^2) \\
         \end{array}
           \right),
\end{equation}
where $A=-\frac{3}{2}iC_A$, $B=\frac{3}{4}C_A$, and $M=C_A-C_B$. The extra minus sign of $H(k)$ has no effect on the energy spectrums, because the four bands are symmetrical with respect to the zero energy. However, the opposite sign of parameter $M$ leads to the helical edge states emerging in the opposite parameter regions for the two edge geometries.
\end{widetext}


\begin{thebibliography}{99}

\bibitem {Hasan2010}M. Z. Hasan and C. L. Kane, Colloquium: Topological insulators, \href{https://doi.org/10.1103/RevModPhys.82.3045}{Rev. Mod. Phys. \textbf{82}, 3045 (2010)}.

\bibitem {Qi2011}X.-L. Qi and S.-C. Zhang, Topological insulators and superconductors, \href{https://doi.org/10.1103/RevModPhys.83.1057}{Rev. Mod. Phys. \textbf{83}, 1057 (2011)}.

\bibitem{Kane2005}C. L. Kane and E. J. Mele, Quantum Spin Hall Effect in Graphene, \href{https://doi.org/10.1103/PhysRevLett.95.226801}{Phys. Rev. Lett. \textbf{95}, 226801 (2005).}

\bibitem{Kane20052}C. L. Kane and E. J. Mele, $Z_2$ Topological Order and the Quantum Spin Hall Effect, \href{https://doi.org/10.1103/PhysRevLett.95.146802}{Phys. Rev. Lett. \textbf{95}, 146802 (2005).}

\bibitem{Bernevig2006}B. A. Bernevig, T. L. Hughes, and S.-C. Zhang, Quantum Spin Hall Effect and Topological Phase Transition in HgTe Quantum Wells, \href{https://doi.org/10.1126/science.1133734}{Science \textbf{314}, 1757 (2006)}.

\bibitem{Konig2007}M. K\"{o}nig, S. Wiedmann, C. Br\"{u}ne, A. Roth, H. Buhmann, L. W. Molenkamp, X.-L. Qi, and S.-C. Zhang, Quantum Spin Hall Insulator State in HgTe Quantum Wells, \href{https://doi.org/10.1126/science.1148047 }{Science \textbf{318}, 766 (2007)}.

\bibitem{Konig2013}M. K\"{o}nig, M. Baenninger, A. G. F. Garcia, N. Harjee, B. L. Pruitt, C. Ames, P. Leubner, C. Br\"{u}ne, H. Buhmann, L. W. Molenkamp, and D. Goldhaber-Gordon, Spatially Resolved Study of Backscattering in the Quantum Spin Hall State, \href{https://doi.org/10.1103/PhysRevX.3.021003}{Phys. Rev. X \textbf{3}, 021003 (2013).}

\bibitem{Roushan2009} P. Roushan, J. Seo, C. V. Parker, Y. S. Hor, D. Hsieh, D. Qian, A. Richardella, M. Z. Hasan, R. J. Cava, and A. Yazdani, Topological surface states protected from backscattering by chiral spin texture, \href{https://doi.org/10.1038/nature08308}{Nature (London) \textbf{460}, 1106 (2009)}.

\bibitem{Chen2018}H. Chen, H. Nassar, A. N. Norris, G. K. Hu, and G. L. Huang, Elastic quantum spin Hall effect in kagome lattices, \href{https://doi.org/10.1103/PhysRevB.98.094302}{Phys. Rev. B \textbf{98}, 094302 (2018)}.

\bibitem{Roth2009}A. Roth, C. Br\"{u}ne, H. Buhmann, L. W. Molenkamp, J. Maciejko, X.-L. Qi, and S.-C. Zhang, Nonlocal Transport in the Quantum Spin Hall State, \href{https://doi.org/10.1126/science.1174736}{Science \textbf{325}, 294 (2009)}.

\bibitem{Brune2012} C. Br\"{u}ne, A. Roth, H. Buhmann, E. M. Hankiewicz, L. W. Molenkamp, J. Maciejko, X.-L. Qi, and S.-C. Zhang, Spin polarization of the quantum spin Hall edge states, \href{https://doi.org/10.1038/nphys2322}{Nat. Phys. \textbf{8}, 485 (2012).}

\bibitem{Hart2014} S. Hart, H. Ren, T. Wagner, P. Leubner, M. M\"{u}hlbauer, C. Br\"{u}ne, H. Buhmann, L. W. Molenkamp, and A. Yacoby, Induced superconductivity in the quantum spin Hall edge, \href{https://doi.org/10.1038/nphys3036}{Nat. Phys. \textbf{10}, 638 (2014).}

\bibitem{Wu2018}S. Wu, V. Fatemi, Q. D. Gibson, K. Watanabe, T. Taniguchi, R. J. Cava, and P. Jarillo-Herrero, Observation of the quantum spin Hall effect up to 100 kelvin in a monolayer crystal, \href{https://doi.org/10.1126/science.aan6003}{Science \textbf{359}, 76 (2018)}.

\bibitem{Freeney2020}S. E. Freeney, J. J. van den Broeke, A. J. J. Harsveld van der Veen, I. Swart, and C. M. Smith, Edge-Dependent Topology in Kekul\'{e} Lattices, \href{https://doi.org/10.1103/PhysRevLett.124.236404}{Phys. Rev. Lett. \textbf{124}, 236404 (2020).}

\bibitem{Fu2011}L. Fu, Topological Crystalline Insulators, \href{https://doi.org/10.1103/PhysRevLett.106.106802}{Phys. Rev. Lett. \textbf{106}, 106802 (2011).}

\bibitem{Slager2013}R.-J. Slager, A. Mesaros, V. Juri\v{c}i\'{c}, and J. Zaanen, The space group classification of topological band-insulators, \href{https://doi.org/10.1038/nphys2513}{Nat. Phys. \textbf{9}, 98 (2013).}

\bibitem{Kariyado2017} T. Kariyado and X. Hu, Topological States Characterized
by Mirror Winding Numbers in Graphene with Bond Modulation, \href{https://doi.org/10.1038/s41598-017-16334-0}{Sci. Rep. \textbf{7}, 16515 (2017)}.

\bibitem{Cao2017}T. Cao, F. Zhao, and S. G. Louie, Topological Phases in Graphene Nanoribbons: Junction States, Spin Centers,
and Quantum Spin Chains, \href{https://doi.org/10.1103/PhysRevLett.119.076401}{Phys. Rev. Lett. \textbf{119}, 076401 (2017)}.

\bibitem{LeeNL2018} Y.-L. Lee, F. Zhao, T. Cao, J. Ihm, and S. G. Louie, Topological
Phases in Cove-Edged and Chevron Graphene Nanoribbons: Geometric Structures, $\mathbb{Z}_2$ Invariants, and Junction States,     \href{https://doi.org/10.1021/acs.nanolett.8b03416}{Nano Lett. \textbf{18}, 7247 (2018).}

\bibitem{Lee2018}C. H. Lee, S. Imhof, C. Berger, F. Bayer, J. Brehm, L. W. Molenkamp, T. Kiessling, and R. Thomale, Topolectrical Circuits, \href{https://doi.org/10.1038/s42005-018-0035-2}{Comm. Phys. \textbf{1}, 39 (2018)}.

\bibitem{Imhof2018}S. Imhof, C. Berger, F. Bayer, J. Brehm, L. W. Molenkamp, T. Kiessling, F. Schindler, C. H. Lee, M. Greiter, T. Neupert, and R. Thomale, Topolectrical-circuit realization of topological corner modes, \href{https://doi.org/10.1038/s41567-018-0246-1}{Nat. Phys. \textbf{14}, 925 (2018).}

\bibitem{Hofmann2019}T. Hofmann, T. Helbig, C. H. Lee, M. Greiter, and R. Thomale, Chiral Voltage Propagation and Calibration in a Topolectrical Chern Circuit, \href{https://doi.org/10.1103/PhysRevLett.122.247702}{Phys. Rev. Lett. \textbf{122}, 247702 (2019)}.

\bibitem{Zhu2019} W. Zhu, Y. Long, H. Chen, and J. Ren, Quantum valley Hall effects and spin-valley locking in topological Kane-Mele circuit networks, \href{https://doi.org/10.1103/PhysRevB.99.115410}{Phys. Rev. B \textbf{99}, 115410 (2019)}.

\bibitem{Lu2019} Y. Lu, N. Jia, L. Su, C. Owens, G. Juzeli\={u}nas, D. I. Schuster, and J. Simon, Probing the Berry curvature and Fermi arcs of a Weyl circuit, \href{https://doi.org/10.1103/PhysRevB.99.020302}{Phys. Rev. B \textbf{99}, 020302(R) (2019)}.

\bibitem{Yyt2020} Y. Yang, D. Zhu, Z. H. Hang, Y. D. Chong, Observation of antichiral edge states in a circuit lattice, \href{https://arxiv.org/abs/2008.10161}{arXiv:2008.10161}.

\bibitem{Yang2020}H. Yang, Z.-X. Li, Y. Liu, Y. Cao, and P. Yan, Observation of symmetry-protected zero modes in topolectrical circuits, \href{https://doi.org/10.1103/PhysRevResearch.2.022028}{Phys. Rev. Research \textbf{2}, 022028(R) (2020)}.

\bibitem{Song2020}L. Song, H. Yang, Y. Cao, and P. Yan, Realization of the square-root higher-order topological insulator in electric circuits, \href{https://doi.org/10.1021/acs.nanolett.0c03049}{Nano Lett. \textbf{20}, 7566 (2020).}

\bibitem{Ezawa2020}M. Ezawa, Braiding of Majorana-like corner states in electric circuits and its non-Hermitian generalization, \href{https://doi.org/10.1103/PhysRevB.100.045407}{Phys. Rev. B \textbf{100}, 045407 (2020)}.

\bibitem{Ezawa20202}M. Ezawa, Electric circuits for non-Hermitian Chern insulators, \href{https://doi.org/10.1103/PhysRevB.100.081401}{Phys. Rev. B \textbf{100}, 081401(R) (2019)}.

\bibitem{SM}See Supplemental Material at http://link.aps.org/ supplemental/ for the form of the circuit Laplacian (Sec. I), the derivation of the mirror winding number (Sec. II), and the mapping to the BHZ model (Sec. III), which includes Refs. \cite{Bernevig2006,Kariyado2017,Yang2020s}.

\bibitem{Yang2020s}Y. Yang, Z. Jia, Y. Wu, Z.-H. Hang, H. Jiang, and X. C. Xie, Gapped topological kink states and topological corner states in graphene, \href{https://doi.org/10.1016/j.scib.2020.01.024}{Sci. Bull. \textbf{65}, 531 (2020)}.

\bibitem{LTspice}LTspice, \href{https://www.linear.com/LTspice}{www.linear.com/LTspice}.

\bibitem{Tamm1932} I. Tamm, \"{U}ber eine m\"{o}gliche Art der Elektronenbindung an Kristalloberfl\"{a}chen, {Phys. Z. Sowjetunion \textbf{76}, 849 (1932)}.

\bibitem{Shockley1939} W. Shockley, On the surface states associated with a periodic potential, \href{https://doi.org/10.1103/PhysRev.56.317}{Phys. Rev. \textbf{56}, 317 (1939)}.

\bibitem{Zhang2008} Y. Zhang, J.-P. Hu, B. A. Bernevig, X. R. Wang, X. C. Xie, and W. M. Liu, Quantum blockade and loop currents in graphene with topological defects, \href{https://doi.org/10.1103/PhysRevB.78.155413}{Phys. Rev. B \textbf{78}, 155413 (2008)}.

\bibitem{Zhang2009} Y. Zhang, J.-P. Hu, B. A. Bernevig, X. R. Wang, X. C. Xie, and W. M. Liu, Localization and the Kosterlitz-Thouless Transition in Disordered Graphene, \href{https://doi.org/10.1103/PhysRevLett.102.106401}{Phys. Rev. Lett. \textbf{102}, 106401 (2009)}.

\bibitem{Wu2016} L.-H. Wu and X. Hu, Topological Properties of Electrons
in Honeycomb Lattice with Detuned Hopping Energy, \href{https://doi.org/10.1038/srep24347}{Sci. Rep. \textbf{6}, 24347 (2016)}.

\bibitem{Fujita1996}M. Fujita, K. Wakabayashi, K. Nakada, and K. Kusakabe, Peculiar localized state at zigzag graphite edge,  \href{https://doi.org/10.1143/JPSJ.65.1920}{J. Phys. Soc. Jpn. \textbf{65}, 1920 (1996)}.
\end{thebibliography}
\end{document}